\documentclass[aps,prd,twocolumn,superscriptaddress,showpacs]{revtex4}
\usepackage[dvips]{graphicx,color}

\begin{document}



\title{Ultra-stable performance of an underground-based
laser interferometer observatory for gravitational waves}


\author{Shuichi Sato} 
\email[]{sato.shuichi@nao.ac.jp} 
\affiliation{National Astronomical Observatory of Japan, 
2-21-1 Osawa, Mitaka, Tokyo, 181-8588, Japan.}
\author{Shinji Miyoki} \affiliation{Institute for Cosmic Ray Research, The University of Tokyo, 
5-1-5, Kashiwanoha, Kashiwa, Chiba 277-8582, Japan.}
\author{Souichi Telada} \affiliation{National Institute of Advanced Industrial Science and Technology, 
1-1-1 Umezono, Tsukuba, Ibaraki, 305-8563, Japan.}
\author{Daisuke Tatsumi} \affiliation{National Astronomical Observatory of Japan, 
2-21-1 Osawa, Mitaka, Tokyo, 181-8588, Japan.}
\author{Akito Araya} \affiliation{Earthquake Research Institute, The University of Tokyo, 
1-1-1, Yayoi, Bunkyo-ku, Tokyo, 113-0032, Japan.}
\author{Masatake Ohashi} \affiliation{Institute for Cosmic Ray Research, The University of Tokyo, 
5-1-5, Kashiwanoha, Kashiwa, Chiba 277-8582, Japan.}
\author{Yoji Totsuka} \altaffiliation{High Energy Accelerator Research Organization (KEK), 
1-1 Oho, Tsukuba, Ibaraki, 305-0801, Japan}
\affiliation{Kamioka Observatory, Institute for Cosmic Ray Research, 
The University of Tokyo, Higashi-Mozumi, Kamioka, Yoshiki-gun, Gifu, 506-1205, Japan.}
\author{Mitsuhiro Fukushima} \affiliation{National Astronomical Observatory of Japan, 
2-21-1 Osawa, Mitaka, Tokyo, 181-8588, Japan.}
\author{Masa-Katsu Fujimoto} \affiliation{National Astronomical Observatory of Japan, 
2-21-1 Osawa, Mitaka, Tokyo, 181-8588, Japan.}

\collaboration{The LISM collaboration}

\date{2004/01/28}
\def\x{\!\times\!}

\begin{abstract}
In order to detect the rare astrophysical events that generate gravitational wave (GW) 
radiation, sufficient stability is required for GW antennas to allow long-term observation.
In practice, seismic excitation is one of the most common disturbances effecting stable 
operation of suspended-mirror laser interferometers.
A straightforward means to allow more stable operation is therefore to locate the 
antenna, the ``observatory'', at a ``quiet'' site. 
A laser interferometer gravitational wave antenna with a baseline length of 20\,m (LISM) 
was developed at a site 1000\,m underground, near Kamioka, Japan.
This project was a unique demonstration of a prototype laser interferometer for gravitational 
wave observation located underground.
The extremely stable environment is the prime motivation for going underground.
In this paper, the demonstrated ultra-stable operation of the interferometer and a 
well-maintained antenna sensitivity are reported.
\end{abstract}

\maketitle


\section{Introduction}

First-generation ground-based gravitational wave antennas (LIGO-I~\cite{ligo1, ligo2}, 
VIRGO~\cite{virgo1, virgo2}, GEO\,600~\cite{geo1, geo2, geo3}, 
TAMA\,300~\cite{tama1, tama2, tama3}) are 
expected to come on-line early in this decade as a global network
searching for astrophysical gravitational wave radiation.
At present, some of the detectors are already operating intermittently,
hoping to observe the spacetime strain of the universe.

The aim of these international projects is to directly detect gravitational radiation,
faint ripples in the spacetime fabric.
There are several kinds of expected astrophysical sources, including chirping 
gravitational waves from inspiraling compact star binaries, burst signals from 
supernovae explosions, and the stochastic background radiation.
The expected event rate of these sources is, however, quite low even if the 
uncertainty of the population estimate~\cite{eventrate1, eventrate2, eventrate3} 
is taken into account,
so, to avoid missing these rare and faint signals, stable operation of the 
detector, keeping the duty cycle and the detector sensitivity high and also keeping 
the data quality high, are indispensable requirements for a gravitational wave 
observatory.
In general, the technologies used in a laser interferometer are based on an 
ultra-high precision measurement pursuing extremely high sensitivity, 
so the instruments are very sensitive to almost all kinds of noise, disturbances 
and drifts.
The noise source that most commonly disturbs stable operation of suspended-mirror 
laser interferometers is seismic excitation.
The most promising solution for this problem is to avoid the source
of these disturbances by selecting a quiet environment for a detector site.

The goal of this project (LISM, Laser Interferometer gravitational-wave Small 
observatory in a Mine) is to demonstrate stable operation of the laser 
interferometer and to obtain high quality data for searching for gravitational waves 
at a well-suited observatory site.

The 20-m baseline laser interferometer was originally developed for various prototyping experiments~\cite{20m1,20m2,20m3,20m4} at the campus 
of The National Astronomical Observatory of Japan, in Mitaka, a suburb of Tokyo 
from 1991 to 1998.
In 1999, it was moved to the Kamioka mine in order to perform long-term, stable 
observations as LISM.
In this paper, the merit of going underground and the demonstrated stable 
operation of the antenna are reported.
The results of data analysis and a GW search will appear as a separate article~\cite{coincidence}.

\section{The KAMIOKA site}
Kamioka is in a mountainous area, about 220\,km west of Tokyo. 
The observatory site is inside a mountain.
The laboratory facility was built 1000\,m underground
beneath the top of the mountain, utilizing some of the tunnel network
that was originally developed for commercial mining activity.
This site is also well-known as the site for the Super-Kamiokande nucleon
decay experiment~\cite{sk} and other cosmic ray experiments,
which all need a low-background environment deep underground.
This site is the most probable candidate
for the future Japanese full-scale gravitational wave antenna project,
LCGT~\cite{lcgt1,lcgt2}.

The most decisive reason why Kamioka was selected as the detector site for LISM 
was that the seismic noise level in the underground facility is extremely low with 
few artificial seismic excitations.
The quiet environment there is an overwhelming benefit for a  suspended-mass laser interferometer because the stability of the system depends 
largely on the lack of seismic and other environmental disturbances.
For the laser interferometer once technical noise are suppressed, the spectral sensitivity 
at the lowest frequency 
region (typically below a few 10\,Hz) is expected to be limited by the seismic noise 
even after the attenuation by the vibration isolation systems,
so low-level seismic noise is quite important in this frequency region.
In addition, slow motion of the ground below 1\,Hz, including micro-seismic noise,
also plays an important role for stable operation of laser interferometer.
The rms value of the seismic displacement in this frequency region
affects lock acquisition, stability, and robustness of the lock.
This is because the suspension systems, which have an eigenfrequency of about 1\,Hz,
cannot be expected to have sufficient isolation at and below that frequency.

\begin{figure}
\includegraphics[width=1.0\linewidth]{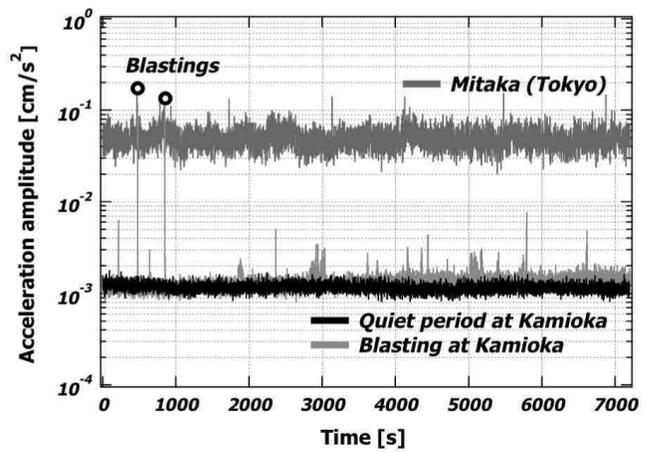}%
\caption{\label{fig1} The acceleration amplitude due to the seismic motion measured with 
accelerometers are shown as a function of time. The acceleration of the Tokyo 
site is typically greater than that of the Kamioka site by about two orders of magnitude. 
Two noise spikes around 500 and 900\,sec are caused by blasting activity near 
the site. These noise spikes have a level almost comparable to the usual acceleration level of Tokyo.}
\end{figure}
\begin{figure}
\includegraphics[width=1.0\linewidth]{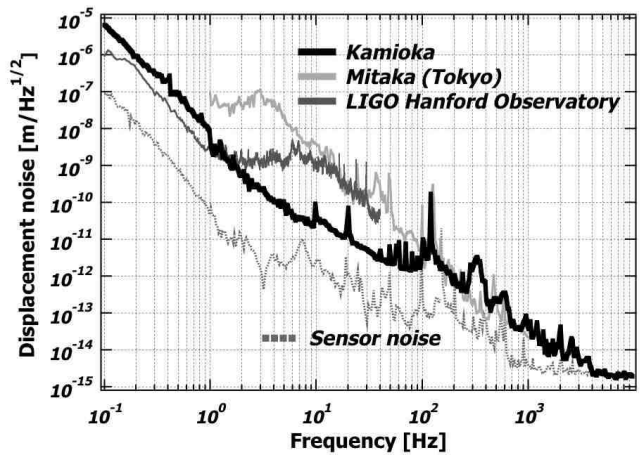}%
\caption{\label{fig2} The seismic noise of the site in horizontal direction is shown in 
displacement as a function of the fourier frequency. The difference in the low frequency 
region below 100\,Hz clearly shows the merit of the Kamioka underground site. 
The seismic displacement noise of LIGO Hanford Observatory (LHO),  
measured inside one of the station buildings\cite{LHO}, is shown as a reference.  
 The frequency range of the LHO spectrum was limited by the bandwidth of  
the seismometer sensor.}
\end{figure}

The typical seismic acceleration is plotted as a function of time at the Kamioka site 
in comparison with that of Tokyo in Fig.~\ref{fig1}.
The rms value of acceleration at the LISM site is about 100 times
smaller than that at the Tokyo site, which is almost comparable
with the transient excitations generated by blasting activities of mining nearby.
The displacement noise power spectrum of seismic motion is shown in Fig.~\ref{fig2}.
The seismic motion is smaller than that of Tokyo by two to three orders
of magnitude in the low frequency region.
The fact that seismic motion is extremely small at lower frequencies
(below 1\,Hz typically) is due to the fact that Kamioka is located in a relatively
quiet region of the Japan islands while the low seismic noise at observational 
frequencies is due to the fact that the laboratory site is deep underground,
1000\,m from surface, so the seismic motion is strongly damped in the high frequency region.

\begin{figure}
\includegraphics[width=1.0\linewidth]{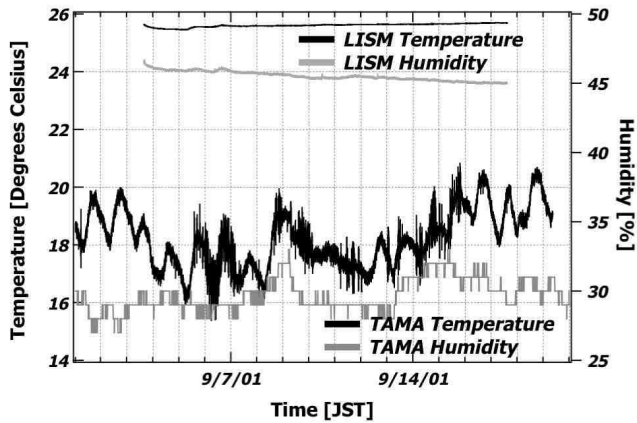}%
\caption{\label{fig3} The temperature and humidity variation in the LISM laboratory 
in comparison with that in the TAMA\,300 site. The variations of the LISM site were almost 
flat, showing 0.01\,degree/day and 0.08\,\%/day for temperature and humidity variation 
respectively. In addition, there were no apparent daily variation and the trend over several 
days caused by the change of the wheather, which was clearly seen in the TAMA\,300 site 
data even though air conditioning is used.}
\end{figure}

In addition, there is a much smaller temperature variation in the underground cave
than at the surface, as shown in Fig.~\ref{fig3}.
As the variation was so small, no temperature control system was employed for the laboratory 
containing the LISM antenna.
This was better than having an active temperature control system.
The temperature stability at the site was considerably
better than that at the surface-based TAMA\,300 site, which is air conditioned.
The temperature variation (drift) in the laboratory was about
$0.01^\circ\,$C each day in the absence of operators.
Human presence was the most significant heat and moisture source in this environment.
Owing to the excellent temperature stability, the drift of the control loops,
the fluctuations in the 
frequency of laser light, the beam pointing, the alignments of optics,
and the variation of other parameters, were expected to be sufficiently suppressed.
Another merit of the underground site should also be mentioned.
There is less influence from bad weather, winds, high wave,
typhoon and so on, which often disrupt the operation of the TAMA\,300 interferometer.

In view of these aspects, the underground environment at Kamioka
is expected to be very suitable for stable long-term operation
of a laser interferometer as a gravitational wave observatory.

\section{Interferometer configuration}
\subsection{Optical configuration}
\begin{figure}
\includegraphics[width=1.0\linewidth]{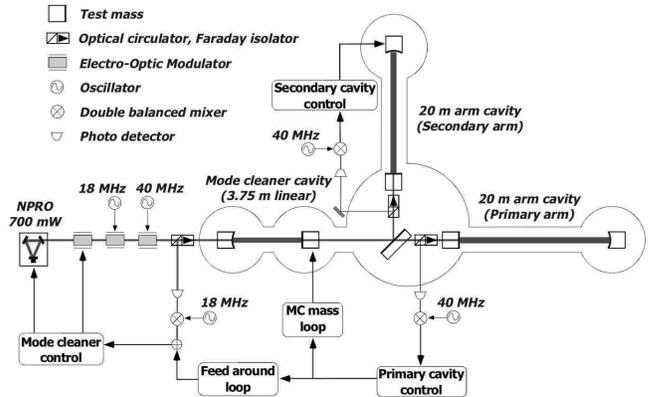}%
\caption{\label{fig4} The schematic of the optical layout and control scheme of the LISM 
interferometer. The main interferometer employs a so called Locked Fabry-Perot 
system, with MISER as a laser source and a linear-type mode cleaner optical cavity.
The laser frequency was stabilized by using the mode cleaner and primary 
cavity as a frequency reference. The multi-stage stabilization scheme was used for 
this purpose with a bandwidth over 1.2\,MHz. The output signal that should contain 
Gravitational signals was extracted from the secondary control loop.} 
\end{figure}

The optical layout of LISM shown n Fig.~\ref{fig4} is based on a Michelson type interferometer whose arms 
contain 20\,m Fabry-Perot optical cavities to enhance the effect of gravitational waves, 
a so-called locked Fabry-Perot interferometer\cite{lfp}.
The frequency of the laser was stabilized by using the primary cavity as a frequency reference.
In other words, the laser frequency was a measure of the spacetime curvature along the 
direction of the primary arm.
By measuring the variation of the secondary cavity length,
which is a reference now of the curvature of spacetime along the perpendicular arm 
cavity direction, it is possible to sense the
differential strain generated by a gravitational wave.

Both Fabry-Perot arm cavities had a finesse of about 25\,000,
which corresponds to a cavity pole frequency of 150\,Hz.
Despite the short arm length, the photon storage time was increased
to 1\,msec by the high finesse.
A commercial Nd:YAG laser running at 1064\,nm, MISER (Light Wave Electronics Co, Ltd),
was used as a light source which yielded an output power of 700\,mW.
The interferometer is equipped with another suspended Fabry-Perot cavity,
the so-called mode cleaner (MC), in front of the main interferometer.
It is quite important for the spatial filtering of the incoming laser light,
and as a first stage reference for the laser frequency stabilization.
The length of the mode cleaner was chosen to be 3.75\,m to have
a free spectral range (FSR) of 40\,MHz,
which enables the 40\,MHz RF-phase modulation
sidebands to pass through it (sideband transmission).

Test masses for the interferometer are made of monolithic fused silica substrates,
whose dimensions are 50\,mm in diameter and 60\,mm in thickness.
The surfaces are super-polished and coated
with high quality dielectric multi-layer films by ion beam sputtering.
The resulting mirrors had an intensity reflection of 0.999875
and a total loss of 27\,ppm~\cite{loss}.
The test masses of the main interferometer and the mode cleaner mirrors
were suspended as double stage pendulums in order to isolate them from seismic noise.
The pendulum resonances at the eigen frequencies (pendulum modes) were suppressed
by using eddy current damping with permanent magnets
attached to the magnet holder surrounding intermediate masses.
For control, four small permanent magnets were glued onto the back of each mirror.
Together with coils fixed on a pendulum cage, the coil-magnet actuators control
the mirror position along the beam axis.

The main interferometer and the mode cleaner are housed
in vacuum chambers connected by 200\,mm diameter vacuum tubes.
During the observation, the vacuum was $10^{-4}$\,Pa and was maintained by 
two ion pumps, as they are free from mechanical vibration.

In order to control the length degrees of freedom of the cavities,
a Pound-Drever-Hall technique was used~\cite{pdh}.
The error signal of the primary cavity was fed back to the laser frequency
via both the MC cavity and a feed-around path~\cite{kawamura}.
The control bandwidth of the MC loop was increased to 1.2\,MHz,
having a loop gain of 80\,dB at 1\,kHz, and that of the primary-cavity loop
was 200\,kHz, with a loop gain of 50\,dB at 1\,kHz.
Owing to this wide-band, high-gain frequency stabilization control,
frequency noise was sufficiently suppressed so as not to contribute
to the resulting noise curve, even though the common mode rejection 
by optical recombination at the beam splitter, as conventionally used, 
was not employed here.
The information of the length of the secondary cavity,
which should contain the gravitational wave signals,
was taken from the feedback of the control loop and recorded.

\subsection{Automatic lock acquisition system}

In order to perform long-term operation of the laser interferometer,
remote control and/or automated lock maintenance systems
are required for a gravitational wave observatory.
This reduces the work load for the operators, and also minimizes
the disturbances caused by the work of the operations.
For the LISM interferometer, a stand-alone automation system was developed.
On detecting the loss of lock, the lock-acquisition sequence
was automatically performed to regain the operational mode.
Owing to this system, the operator could simply monitor
the interferometer condition from outside the mine.
These systems are based on hardware control using a TTL digital signal.
The state of the interferometer was judged by monitoring
the optical power of the transmission and/or
the reflection of the Fabry-Perot cavities.
This information was communicated as a TTL signal, to engage,
switch, and change the characteristics of the analog control loops
via appropriate digital circuitry that provides sequential timing signals.
One aspect which is different from that at the Tokyo site
is that the motion of the suspended mirrors is so small
that it takes a long time for the cavities to pass
through one free spectral range (FSR).
In other words, as the Fabry-Perot cavities have a
resonance flash so infrequently, they can not acquire lock
in a reasonable time by themselves.
Without an automatic lock acquisition system, i.e.\ if the control systems
are left waiting for self re-lock, it typically took a day to re-acquire
lock and become operational after an unlock of the interferometer.
This will be mentioned again and justified later in the result section.

For quick lock acquisition, the wave-length of the laser light
or the length of the cavities were actively swept
by injecting sweeping signals to the feedback points
of the control loops until the cavities acquired lock.
The sweeping was performed softly and slowly enough
that the control system could acquire lock even with the high-finesse arm cavities.
Once one of the cavities was locked, the sequential switching processes were initiated.
The main functions of these processes were the change of the control servo
(servo gain increase, DC gain booster, etc.),
optical power increase (reduced power was used for lock acquisition),
and switching of the circuits for sweeping-signal injection.
Using this system, the whole interferometer became operational
within 100\,sec typically.
Once the systems were properly tuned,
this automated lock acquisition system worked very reliably,
and the failure rate of the re-lock acquisition sequence
was very low, which strongly contributed to the excellent duty cycle.

\section{Interferometer sensitivity and noise budget}

\begin{figure}
\includegraphics[width=1.0\linewidth]{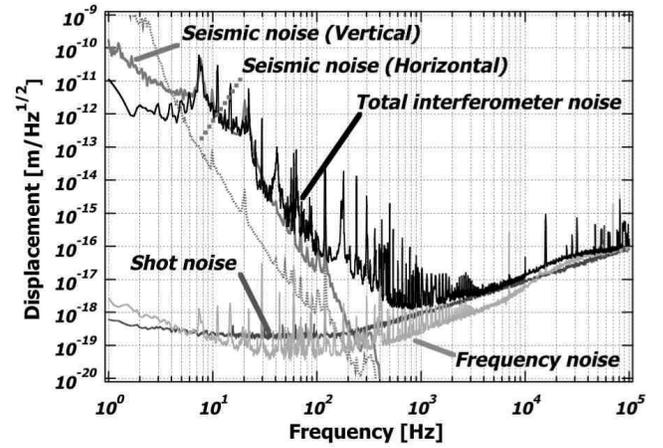}%
\caption{\label{fig5} The noise equivalent detector sensitivity is shown in 
displacement together with identified noise sources that were limiting  
the sensitivity. The frequency noise was surpressed well and optical shot noise 
was the dominant contribution in the higher frequency region above 1\,kHz. 
In the lower frequency region below 100\,Hz, seismic noise in the vertical 
direction which couples through the suspension system exceeded direct 
horizontal seismic noise and had a significant influence on the detector sensitivity.} 
\end{figure}

The noise equivalent spectral sensitivity curve of the LISM interferometer
is shown in Fig.~\ref{fig5}, together with identified noise sources.
After much effort to reduce noise, the detector achieved 
a floor sensitivity of $1.3 \x 10^{-18}\rm\, m/\sqrt{Hz}$
around 800\,Hz in displacement,
which corresponds to $6\x10^{-20}/\sqrt{\rm Hz}$ in strain.
Noise sources that limited the detector sensitivity
were to a large part identified in a wide frequency range.

For the high frequency range above 1\,kHz, photon shot noise was dominant,
corresponding to the input light power to the secondary cavity of 35\,mW.
The arm cavities are designed to be critically coupled so that only the rf modulation 
sidebands are reflected to the detector.
In this regime the shot noise limited sensitivity is maximized in the limit where the rf 
sidebands power goes to zero.
However, there is some amount of carrier reflection due to finite optical losses 
of the optics and mismatch of reflectivity between the two cavity mirrors.
In the regime where the detected power is dominated by the carrier, the shot 
noise limited sensitivity can be improved by increasing the amplitude of the 
rf sidebands until the sidebands power dominates. 
Therefore the rf sidebands power was adjusted to optimize the signal to noise ratio.

For the lower frequencies, below several hundred Hz,
the seismic noise was dominant.
The calculated seismic noise level in the horizontal direction,
using the measured seismic noise spectrum and the calculated
transfer function of the suspension system,
showed no contribution to the interferometer sensitivity curve
in the observational band.
On the other hand, the estimated contribution of the vertical seismic noise,
which can be converted to horizontal motion
via vertical-horizontal cross coupling of the suspension system,
could explain well a resonance structure around a few tens of Hertz.
In the lowest frequency region, below 6\,Hz, the noise curve
of the interferometer is below the expectation
of seismic noise contributions.
This is thought to be an effect of common mode suppression.
The velocity of the elastic wave, which is a function of the elasticity modulus 
and matter density, at this location is rather fast, so the wave length of the 
seismic motion at low frequency is presumed to be comparable to the 20\,m arm length.
Therefore, in this low-frequency region, the two test masses composing the
Fabry-Perot cavity are considered to be moving together, so there is less 
relative displacement between them.
This is more evidence that the ground of the site Kamioka 
underground is stiff and firm, where the primary wave velocity is the order of 
5500\,m/s.
The accumulated displacement between two test masses was calculated
to be of the order of $10^{-10}\,$m, by integrating the noise spectrum
down to 0.1\,Hz, which corresponds to an accumulated rms velocity
on the order of $10^{-9}\,$m/s.
This means that the round trip phase of the Fabry-Perot cavity changes
very slowly, owing largely to the significant common mode rejection
of displacement between the cavity mirrors.
This is consistent with the fact that the antenna took
one night to re-lock by itself if left as is.
The excess noise between these two frequency regions (shot noise limited high frequency 
region and seismic noise limited low frequency region), around 100\,Hz,
was suspected to come from unwanted coupling through the suspension systems.
There are several candidates for the source of noise which coupled through the 
suspension systems, and it is not clear which was the problem, however by 
improving the suspension system this noise was eliminated.
Consequently, all the noise sources contributing to the interferometer
sensitivity were fully identified as shot noise, seismic noise, and suspension oriented noise.

 \section{Observation and results}
Several observational runs were performed from the beginning of 2000,
with an accumulated observation time on the order of 3000 hours.
Of significance, 1000 hours of data were taken in the summer of 2001 in coincidence
with the TAMA\,300 detector.
The report on the development of the coincident analysis method and the results
using LISM and TAMA\,300 data will appear as an independent article elsewhere.
In this article, the achievement of ultra-stable operation
of a gravitational wave antenna and its importance are emphasized.

 \subsection{Tidal motion}

\begin{figure}
\includegraphics[width=1.0\linewidth]{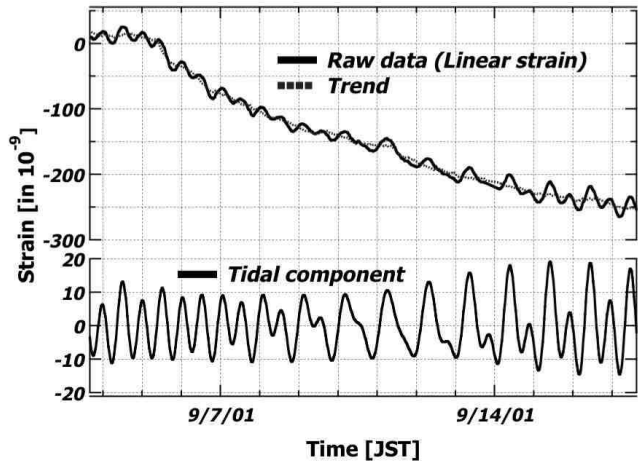}%
\caption{\label{fig6} The strain of the arm cavity relative to the primary 
cavity. Discontinuities of the raw data correspond to the cavity unlock.
The tidal modulation components are clearly seen on the trend.
The geophysical response at the site to the tidal force turned out to be on 
the order of $10^{-8}$.} 
\end{figure}

During the operation of the interferometer, the Fabry-Perot cavities
were controlled in length in order to keep them on resonance.
In principle, the mismatch between cavity length and wavelength
of the laser light is sensed and then fed back to their actuators
via appropriate servo filters so that these feedback signals
cancel the deviations.
In other words, that feedback signal contains the information
for the change of the cavity length together with the change
of the wavelength of the laser light, which can be caused by,
for example, by temperature variation.
%
The PZT actuator of the laser was used for a laser frequency tuning 
in low frequency region, so the feedback signals on the actuator for the 
primary cavity was converted to strain equivalent quantities in Fig.~\ref{fig6}.
These variations include both the change of the cavity length itself
and that of the wavelength of the laser light,
and these two effects cannot be distinguished in principle.
Although there is a large drift over several days in one direction,
a tidal component was clearly seen on it.
This shows that there is less seismic motion and environmental disturbances
(for example temperature variations)
than that which would screen the tidal strain components.
The observed equivalent strain was separated into three components,
namely long-term drift, noise, and tidal effects.
The tidal component was calculated by tidal analysis software,
Baytap-G\cite{baytap}, using the interferometer location ($\rm 137.18^\circ E,36.25^\circ N$)
and the orientation.
By fitting the data, the strain at the Kamioka site caused by tidal motion
of the earth was on the order of $10^{-8}$, which was almost same with calculated value.

\subsection{Operational stability}

\begin{figure}
\includegraphics[width=1.0\linewidth]{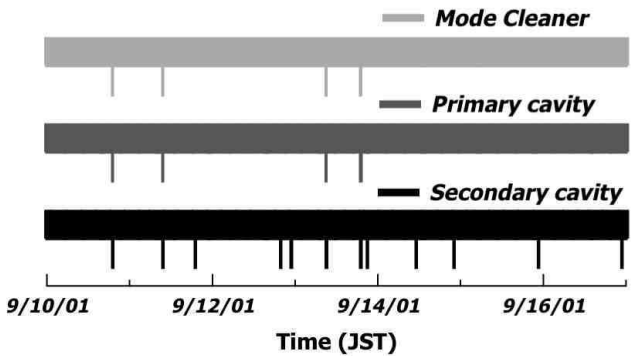}%
\caption{\label{fig7} The operational status of the three cavities of the interferometer 
are shown for a 1 week period. Each bar shows ``operational status'', the lines below 
the bars represent unlocks. There were 12 unlocks during this period, however 
there  was no long unlock which significantly lowered the duty cycle of the antenna. 
The resulting duty cycle was 99.8\,\%. Most of these unlock were due to the impulsive seismic 
noise caused by blasting.} 
\end{figure}

The operational status of the interferometer for one week from the 
two months of observational operation is shown in Fig.~\ref{fig7} 
with Japanese standard time (JST) on the horizontal axis.
Three bars show the lock status of each of the three cavities.
The mode cleaner, the primary, and the secondary cavities.
Lock was lost a total of 12 times during this period,
10 of these 12 times, the cause was blasting activity
of the mining company.
However the interferometer typically became operational again after about 100\,s
(the shortest recovery occurred within a few seconds),
and the resulting accumulated dead (loss) time was only 1440 seconds
during that 7 day period, so the antenna was operational 167.6 hours out of the 168 hours
of total observation time in this period, which corresponds to a duty cycle of 99.8\%.
This is a tribute to the environmental stability of the site, and the automation systems.
The blasting from the mining activities have been terminated in the mountain,
so the frequency of the unlocks in the future observations will be significantly reduced,
and longer stretches of lock are expected.
Actually the LISM detector had a record of 270 hours of continuous operation
in another observation period, when blasting was not being done.

\subsection{Sensitivity stability}

\begin{figure}
\includegraphics[width=1.0\linewidth]{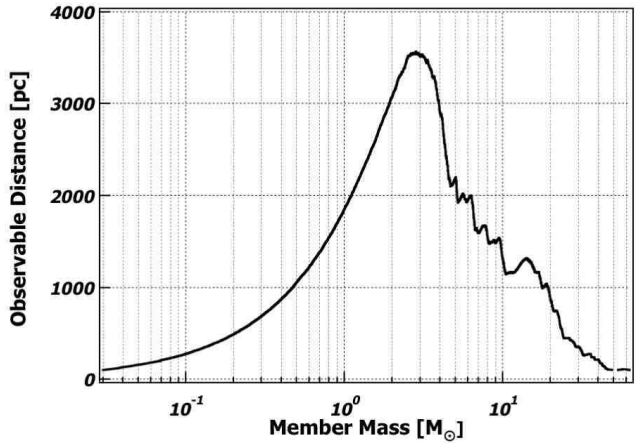}%
\caption{\label{fig8} The expected binary range of the gravitational wave signals 
from equal mass binary inspirals with SNR=10 by matched filtering method. For 
simplicity optimal source direction and polarization are assumed.} 
\end{figure}

\begin{figure}
\includegraphics[width=1.0\linewidth]{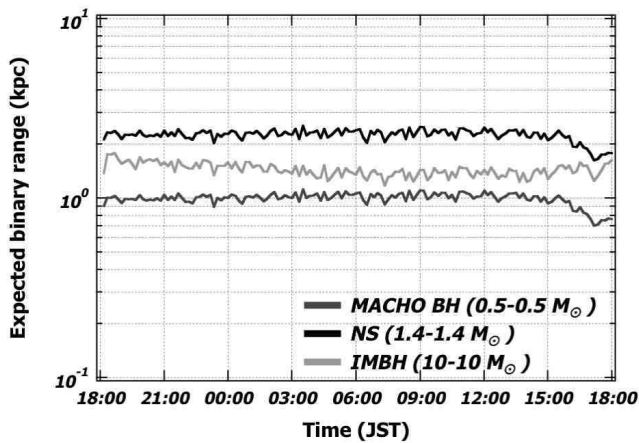}%
\caption{\label{fig9} The stability of the expected binary range for 0.5, 1.4 and 10 
solar mass binary systems for 24 hours. There seemed no significant decrease 
in each binary range, which means that the LISM antenna was stably operated 
to keep initial spectral sensitivity. Another emphasis should be on the fact that 
there was no day-night effect, which is the case at the Tokyo site because of various 
human activities in the daytime.} 
\end{figure}

From the point of view of the observation of gravitational waves,
the stability of the sensitivity and the data quality are all vital issues
for searching for gravitational wave signals.
There may, however, be requirements for the data quality that differ
depending on the target GW signals and the analysis methods.
There also are several ways to evaluate the spectral sensitivity,
for example the spectrogram, which is sometimes used
for a burst event search~\cite{ando}.
Here we adopt the expected binary range of a matched filtering analysis as an 
index to evaluate the spectrum sensitivity quantitatively.
The binary range is defined as follows \cite{tagoshi} using the spectral sensitivity of the detector 
with assumed specific signal-to-noise ratio (SNR) value, 
\begin{eqnarray}
\textrm{Binary Range}&=&4A\left[\int_{0}^{f_{c}}\frac{f^{-7/3}}{S_{n}(f)}{\rm d}f\right]^{1/2} \\
  & & A=\frac{T_{\odot}^{5/6}c}{\textrm{SNR}} \left(\frac{5\mu}{96M_{\odot}}\right)^{1/2}
   \left(\frac{M}{\pi^{2}M_{\odot}}\right)^{1/3}\,.
\end{eqnarray}
Here, $S_{n}(f)$ is the noise power spectrum of the detector, 
$c$ is the speed of light, $\mu=m_{1}m_{2}/M$ is the reduced mass, 
$M=m_{1}+m_{2}$ is the total mass of the binary system,
$T_{\odot}=(G/c^{3})M_{\odot}$ is the solar time 
and  $M_{\odot}$ is the solar mass.
The cutoff frequency was defined as $f_{c}=M_{\odot}/(6^{3/2}\pi MT_{\odot})$ in this calculation.
As the matched filtering analysis is performed over some
Fourier frequency ranges defined by a particular cutoff frequency
depending on the chirp mass, any change of spectral sensitivity
in these frequency regions influences the results of this evaluation.
An expected binary range for gravitational wave radiation from coalescence
of equal mass binary systems is shown in Fig.~\ref{fig8}
as a function of member star mass.
This shows how far the detector can see the GW event with SNR=10.
The detector was, for simplicity, assumed to be optimally oriented
to the gravitational wave source both in direction
and its radiation polarization.
According to this result, LISM has a sensitivity to detect
inspiraling gravitational waves from 1.4--1.4\,$M_{\odot}$
neutron star coalescence events that occurred
3\,kpc away with a SNR of about 10.
The recorded variation of expected binary range for coalescence of 0.5\,$M_{\odot}$,
1.4\,$M_{\odot}$ and 10\,$M_{\odot}$ equal-mass binary systems
is shown in Fig.~\ref{fig9}, for a time stretch of about 24 hours of observation.
There was no significant sensitivity decrease during this period,
and the important result was that there was no day-night effect
on the trend as was strongly experienced at the  Tokyo site.
This shows that the antenna was kept operable in day time
as well as in night time, which means that human activities
in the day time had almost no effect on the antenna at this site.
The distribution of the binary range value for some stable period
shows good gaussianity, with a relatively small standard deviation.
If we define a `sensitive-operation' duty cycle,
which allows a 3\,dB decrease of the SNR from the center value,
instead of the `in-lock operation' duty cycle,
it becomes about 90\% for each mass binary systems.

\section{Conclusion}
The prototype laser interferometer gravitational wave antenna
LISM
with an arm length of 20\,m was moved into a deep underground
laboratory, a first for a gravitational wave observatories, 
and a long-term observational run was performed.
The goal of this project was to confirm that the laser interferometer
can operate stably and provide high quality data,
making the most of the stable environment in the underground laboratory.
The interferometer sensitivity achieved was
$1\x10^{-18}\rm\,m/\sqrt{Hz}$
around 800\,Hz in displacement,
which corresponds to $5\x10^{-20}\rm/\sqrt{Hz}$ in strain.
At higher frequencies, above that floor sensitivity,
LISM was quantum noise limited by shot noise. 
With this sensitivity spectrum, LISM can detect
gravitational wave events emitted from coalescence
of 1.4--1.4\,$M_{\odot}$ equal-mass binary systems a few\,kpc
away by matched filtering analysis with a SNR of 10.
The operational duty cycle of the interferometer
exceeded 99.8\% owing to the stable,
low-disturbance environment and the self-recovering automation systems.
If we adopt the expected SNR of the matched filtering analysis
as an index for stability of the interferometer sensitivity,
this SNR duty cycle was about 90\%.
This means that the interferometer sensitivity was kept 
within a 3\,dB window for 90\% of the observational period.
According to these results, we conclude that
the laser interferometer gravitational wave antenna was operated stably
enough for a long term observational run, and the underground environment
is suitable as a gravitational wave antenna site.

\begin{acknowledgments}
The authors are grateful to the LISM collaboration for their assistance and also 
would like to thank A.~R\"udiger for valuable discussion.
The authors also would like to thank R.~Savage for providing the LHO seismic 
noise spectrum data.
This research was partially supported by the Ministry of Education, Science, Sports and Culture, Grant-in-Aid for Scientific Research (A), 11304013, 1999.
\end{acknowledgments}


\bibliography{bibliography}

\begin{thebibliography}{29}
\expandafter\ifx\csname natexlab\endcsname\relax\def\natexlab#1{#1}\fi
\expandafter\ifx\csname bibnamefont\endcsname\relax
  \def\bibnamefont#1{#1}\fi
\expandafter\ifx\csname bibfnamefont\endcsname\relax
  \def\bibfnamefont#1{#1}\fi
\expandafter\ifx\csname citenamefont\endcsname\relax
  \def\citenamefont#1{#1}\fi
\expandafter\ifx\csname url\endcsname\relax
  \def\url#1{\texttt{#1}}\fi
\expandafter\ifx\csname urlprefix\endcsname\relax\def\urlprefix{URL }\fi
\providecommand{\bibinfo}[2]{#2}
\providecommand{\eprint}[2][]{\url{#2}}

\bibitem[{\citenamefont{{A.~Abramovici, W.E.~Althouse, R.W.P.~Drever,
  Y.~G\"ursel, S.~Kawamura, F.J.~Raab, D.~Shoemaker, L.~Sievers, R.E.~Spero,
  K.S.~Thome, R.E.~Vogt, R.~Weiss, S.E.~Whitcomb, M.E.~Zucker}}(1992)}]{ligo1}
\bibinfo{author}{\bibnamefont{{A.~Abramovici, W.E.~Althouse, R.W.P.~Drever,
  Y.~G\"ursel, S.~Kawamura, F.J.~Raab, D.~Shoemaker, L.~Sievers, R.E.~Spero,
  K.S.~Thome, R.E.~Vogt, R.~Weiss, S.E.~Whitcomb, M.E.~Zucker}}},
  \bibinfo{journal}{Science} \textbf{\bibinfo{volume}{256}},
  \bibinfo{pages}{325} (\bibinfo{year}{1992}).

\bibitem[{\citenamefont{{B.~Barish and R.~Weiss}}(1999)}]{ligo2}
\bibinfo{author}{\bibnamefont{{B.~Barish and R.~Weiss}}},
  \bibinfo{journal}{Phys. Today} \textbf{\bibinfo{volume}{52}},
  \bibinfo{pages}{44} (\bibinfo{year}{1999}).

\bibitem[{\citenamefont{{C.~Bradaschia {\it et al.}}}(1990)}]{virgo1}
\bibinfo{author}{\bibnamefont{{C.~Bradaschia {\it et al.}}}},
  \bibinfo{journal}{Nucl. Instrum. Methods Phys. Res. A}
  \textbf{\bibinfo{volume}{289}}, \bibinfo{pages}{518} (\bibinfo{year}{1990}).

\bibitem[{\citenamefont{{F.~Acernese {\it et al.}}}(2002)}]{virgo2}
\bibinfo{author}{\bibnamefont{{F.~Acernese {\it et al.}}}},
  \bibinfo{journal}{Class. Quantum Grav.} \textbf{\bibinfo{volume}{19}},
  \bibinfo{pages}{1421} (\bibinfo{year}{2002}).

\bibitem[{\citenamefont{{K.~Danzmann {\it et al.}}}(1994)}]{geo1}
\bibinfo{author}{\bibnamefont{{K.~Danzmann {\it et al.}}}},
  \bibinfo{journal}{{Internal Rep. MPQ Max-Planck-lnstitut f\"ur Quantenoptik,
  Garching, Germany}} \textbf{\bibinfo{volume}{190}} (\bibinfo{year}{1994}).

\bibitem[{\citenamefont{{K.~Danzmann}}(1994)}]{geo2}
\bibinfo{author}{\bibnamefont{{K.~Danzmann}}}, in
  \emph{\bibinfo{booktitle}{First E. Amaldi Conf. on Gravitational Wave
  Experiments}}, edited by \bibinfo{editor}{\bibnamefont{{E.~Coccia,
  G.~Pizzella, F.~Ronga}}} (\bibinfo{publisher}{World Scientific, Singapore},
  \bibinfo{year}{1994}), pp. \bibinfo{pages}{100--111}.

\bibitem[{\citenamefont{{B.~Willke {\it et al.}}}(2002)}]{geo3}
\bibinfo{author}{\bibnamefont{{B.~Willke {\it et al.}}}},
  \bibinfo{journal}{Class. Quantum Grav.} \textbf{\bibinfo{volume}{19}},
  \bibinfo{pages}{1377} (\bibinfo{year}{2002}).

\bibitem[{\citenamefont{Tsubono}(1994)}]{tama1}
\bibinfo{author}{\bibfnamefont{K.}~\bibnamefont{Tsubono}}, in
  \emph{\bibinfo{booktitle}{First E. Amaldi Conf. on Gravitational Wave
  Experiments}}, edited by \bibinfo{editor}{\bibnamefont{{E.~Coccia,
  G.~Pizzella, F.~Ronga}}} (\bibinfo{publisher}{World Scientific, Singapore},
  \bibinfo{year}{1994}), pp. \bibinfo{pages}{112--114}.

\bibitem[{\citenamefont{{M.~Ando {\it et al.}}}(2002)}]{tama2}
\bibinfo{author}{\bibnamefont{{M.~Ando {\it et al.}}}},
  \bibinfo{journal}{Class. Quantum Grav.} \textbf{\bibinfo{volume}{19}},
  \bibinfo{pages}{1409} (\bibinfo{year}{2002}).

\bibitem[{\citenamefont{{M.~Ando and the TAMA collaboration}}(2001)}]{tama3}
\bibinfo{author}{\bibnamefont{{M.~Ando and the TAMA collaboration}}},
  \bibinfo{journal}{Phys. Rev. Lett.} \textbf{\bibinfo{volume}{86}},
  \bibinfo{pages}{3950} (\bibinfo{year}{2001}).

\bibitem[{\citenamefont{{V.~Kalogera, R.~Narayan, D.N.~Sperge and
  J.H.~Taylor}}(2001)}]{eventrate1}
\bibinfo{author}{\bibnamefont{{V.~Kalogera, R.~Narayan, D.N.~Sperge and
  J.H.~Taylor}}}, \bibinfo{journal}{Astrophys. Journ.}
  \textbf{\bibinfo{volume}{556}}, \bibinfo{pages}{340} (\bibinfo{year}{2001}).

\bibitem[{\citenamefont{{C.~Kim, V.~Kalogera, and
  D.R.~Lorimer}}(2003)}]{eventrate2}
\bibinfo{author}{\bibnamefont{{C.~Kim, V.~Kalogera, and D.R.~Lorimer}}},
  \bibinfo{journal}{Astrophys. Journ.} \textbf{\bibinfo{volume}{584}},
  \bibinfo{pages}{985} (\bibinfo{year}{2003}).

\bibitem[{\citenamefont{{M.~Burgay, N.~DfAmico, A.~Possenti, R.N.~Manchester,
  A.G.~Lyne, B.C.~Joshi, M.A.~McLaughlin, M.~Kramer, J.M.~Sarkissian,
  F.~Camilo, V.~Kalogera, C.~Kim and D.R.~Lorimer}}(2003)}]{eventrate3}
\bibinfo{author}{\bibnamefont{{M.~Burgay, N.~DfAmico, A.~Possenti,
  R.N.~Manchester, A.G.~Lyne, B.C.~Joshi, M.A.~McLaughlin, M.~Kramer,
  J.M.~Sarkissian, F.~Camilo, V.~Kalogera, C.~Kim and D.R.~Lorimer}}},
  \bibinfo{journal}{Nature} \textbf{\bibinfo{volume}{426}},
  \bibinfo{pages}{531} (\bibinfo{year}{2003}).

\bibitem[{\citenamefont{{M.~Ohashi, M.-K.~Fujimoto, T.~Yamazaki, M.~Fukushima,
  A.~Araya, and S.~Telada}}(1996)}]{20m1}
\bibinfo{author}{\bibnamefont{{M.~Ohashi, M.-K.~Fujimoto, T.~Yamazaki,
  M.~Fukushima, A.~Araya, and S.~Telada}}}, in
  \emph{\bibinfo{booktitle}{Proceedings of the Seventh Marcel Grossmann Meeting
  on General Relativity}}, edited by \bibinfo{editor}{\bibnamefont{{Robert T.
  Jantzen, G. Mac Keiser}}} (\bibinfo{publisher}{World Scientific, Singapore},
  \bibinfo{year}{1996}), pp. \bibinfo{pages}{1370--1371}.

\bibitem[{\citenamefont{{A.~Araya, N.~Mio, K.~Tsubono, K.~Suehiro, S.~Telada,
  M.~Ohashi and M.-K.~Fujimoto}}(1997)}]{20m2}
\bibinfo{author}{\bibnamefont{{A.~Araya, N.~Mio, K.~Tsubono, K.~Suehiro,
  S.~Telada, M.~Ohashi and M.-K.~Fujimoto}}}, \bibinfo{journal}{Appl. Opt.}
  \textbf{\bibinfo{volume}{36}}, \bibinfo{pages}{1446} (\bibinfo{year}{1997}).

\bibitem[{\citenamefont{{S.~Telada, K.~Suehiro, S.~Sato. M.~Ohashi,
  M.-K.~Fujimoto and A.~Araya}}(1996)}]{20m3}
\bibinfo{author}{\bibnamefont{{S.~Telada, K.~Suehiro, S.~Sato. M.~Ohashi,
  M.-K.~Fujimoto and A.~Araya}}}, in \emph{\bibinfo{booktitle}{Proc. of the 1st
  TAMA International Workshop on Gravitational Wave Detection}}, edited by
  \bibinfo{editor}{\bibnamefont{{K.~Tsubono, M.-K.~Fujimoto and K.~Kuroda}}}
  (\bibinfo{publisher}{Universal Academy Press, Inc.}, \bibinfo{year}{1996}),
  pp. \bibinfo{pages}{349--351}.

\bibitem[{\citenamefont{{S.~Sato, M.~Ohashi, M.-K.~Fujimoto, M.~Fukushima,
  K.~Waseda, S.~Miyoki, N.~Mavalvala, and H.~Yamamoto}}(2000)}]{20m4}
\bibinfo{author}{\bibnamefont{{S.~Sato, M.~Ohashi, M.-K.~Fujimoto,
  M.~Fukushima, K.~Waseda, S.~Miyoki, N.~Mavalvala, and H.~Yamamoto}}},
  \bibinfo{journal}{Appl. Opt.} \textbf{\bibinfo{volume}{39}},
  \bibinfo{pages}{4616} (\bibinfo{year}{2000}).

\bibitem[{\citenamefont{{H.~Takahashi {\it et al.}}}()}]{coincidence}
\bibinfo{author}{\bibnamefont{{H.~Takahashi {\it et al.}}}}, \bibinfo{note}{{in
  preparation}}.

\bibitem[{\citenamefont{{Y.~Fukuda {\it et al.}}}(1998)}]{sk}
\bibinfo{author}{\bibnamefont{{Y.~Fukuda {\it et al.}}}},
  \bibinfo{journal}{Phys. Rev. Lett.} \textbf{\bibinfo{volume}{81}},
  \bibinfo{pages}{1562} (\bibinfo{year}{1998}).

\bibitem[{\citenamefont{{K.~Kuroda, M.~Ohashi, S.~.~Miyoki, D.~Tatsumi,
  S.~Sato, H.~Ishizuka, M.-K.~Fujimoto, S.~Kawamura, R.~Takahashi, T.~Yamazaki,
  K.~Arai, M.~Fukushima, K.~Waseda, S.~Telada, A.~Ueda}}(1999)}]{lcgt1}
\bibinfo{author}{\bibnamefont{{K.~Kuroda, M.~Ohashi, S.~.~Miyoki, D.~Tatsumi,
  S.~Sato, H.~Ishizuka, M.-K.~Fujimoto, S.~Kawamura, R.~Takahashi, T.~Yamazaki,
  K.~Arai, M.~Fukushima, K.~Waseda, S.~Telada, A.~Ueda}}},
  \bibinfo{journal}{Inter. Jour. of Mod. Phys. D} \textbf{\bibinfo{volume}{8}},
  \bibinfo{pages}{557} (\bibinfo{year}{1999}).

\bibitem[{\citenamefont{{K.~Kuroda {\it et al.}}}(2003)}]{lcgt2}
\bibinfo{author}{\bibnamefont{{K.~Kuroda {\it et al.}}}},
  \bibinfo{journal}{Class. Quantum Grav.} \textbf{\bibinfo{volume}{20}},
  \bibinfo{pages}{S871} (\bibinfo{year}{2003}).

\bibitem[{\citenamefont{{D. Shoemaker {\it et al.}}}(2001)}]{LHO}
\bibinfo{author}{\bibnamefont{{D. Shoemaker {\it et al.}}}},
  \emph{\bibinfo{title}{The ligo observatory environment}}
  (\bibinfo{year}{2001}), \bibinfo{note}{{LIGO technical document
  LIGO-T010074-03-D} {\tt http://www.ligo.caltech.edu/docs/T/T010074-03/
  T010074-03.pdf}}.

\bibitem[{\citenamefont{Zucker}(1991)}]{lfp}
\bibinfo{author}{\bibfnamefont{M.}~\bibnamefont{Zucker}}, in
  \emph{\bibinfo{booktitle}{Proc. 6th Marcel Grossmann Meeting on General
  relativity, Kyoto, Japan}} (\bibinfo{publisher}{World Scientific, Singapore,
  1991}, \bibinfo{year}{1991}).

\bibitem[{\citenamefont{{S.~Sato, S.~Miyoki, M.~Ohashi,
  M.-K.~Fujimoto,T.~Yamazaki, M.~Fukushima, A.~Ueda, K.~Ueda, K.~Watanabe,
  K.~Nakamura, K.~Etoh, N.~Kitajima, K.~Ito and I.~Kataoka}}(1999)}]{loss}
\bibinfo{author}{\bibnamefont{{S.~Sato, S.~Miyoki, M.~Ohashi,
  M.-K.~Fujimoto,T.~Yamazaki, M.~Fukushima, A.~Ueda, K.~Ueda, K.~Watanabe,
  K.~Nakamura, K.~Etoh, N.~Kitajima, K.~Ito and I.~Kataoka}}},
  \bibinfo{journal}{Appl. Opt.} \textbf{\bibinfo{volume}{38}},
  \bibinfo{pages}{2880} (\bibinfo{year}{1999}).

\bibitem[{\citenamefont{{R.W.P.~Drever, J.L.~Hall, F.V.~Kowalski, J.~Hough,
  G.M.~Ford, A.J.~Munley, and H.~Ward}}(1983)}]{pdh}
\bibinfo{author}{\bibnamefont{{R.W.P.~Drever, J.L.~Hall, F.V.~Kowalski,
  J.~Hough, G.M.~Ford, A.J.~Munley, and H.~Ward}}}, \bibinfo{journal}{Appl.
  Phys. B} \textbf{\bibinfo{volume}{31}}, \bibinfo{pages}{97}
  (\bibinfo{year}{1983}).

\bibitem[{\citenamefont{{S.~Kawamura, A.~Abramovici,
  M.E.~Zucker}}(1997)}]{kawamura}
\bibinfo{author}{\bibnamefont{{S.~Kawamura, A.~Abramovici, M.E.~Zucker}}},
  \bibinfo{journal}{Rev. Sci. Instrum.} \textbf{\bibinfo{volume}{68}},
  \bibinfo{pages}{223} (\bibinfo{year}{1997}).

\bibitem[{\citenamefont{{Y.~Tamura, T.~Sato, M.~Ooe and
  M.~Ishigro}}(1991)}]{baytap}
\bibinfo{author}{\bibnamefont{{Y.~Tamura, T.~Sato, M.~Ooe and M.~Ishigro}}},
  \bibinfo{journal}{Geophys. J. Int.} \textbf{\bibinfo{volume}{104}},
  \bibinfo{pages}{507} (\bibinfo{year}{1991}).

\bibitem[{\citenamefont{{M.~Ando}}(2001)}]{ando}
\bibinfo{author}{\bibnamefont{{M.~Ando}}} (\bibinfo{year}{2001}),
  \bibinfo{note}{{Internal document}}.

\bibitem[{\citenamefont{{H.~Tagoshi {\it et al.}}}(2001)}]{tagoshi}
\bibinfo{author}{\bibnamefont{{H.~Tagoshi {\it et al.}}}},
  \bibinfo{journal}{Phys. Rev. D} \textbf{\bibinfo{volume}{63}},
  \bibinfo{pages}{062001} (\bibinfo{year}{2001}).

\end{thebibliography}

\end{document}